\documentclass[aps,prb,letterpaper,amsmath,amssymb,superscriptaddress,reprint]{revtex4-1}
\usepackage{graphicx}

\begin{document}
\title{$GW$ study of the half-metallic Heusler compounds Co$_2$MnSi and Co$_2$FeSi}
\author{Markus Meinert}
\email{meinert@physik.uni-bielefeld.de}
\affiliation{Thin Films and Physics of Nanostructures, Department of Physics, Bielefeld University, D-33501 Bielefeld, Germany}
\author{Christoph Friedrich}
\affiliation{Peter Gr\"unberg Institut and Institute for Advanced Simulation, Forschungszentrum J\"ulich and JARA, 52425 J\"ulich, Germany}
\author{G\"unter Reiss}
\affiliation{Thin Films and Physics of Nanostructures, Department of Physics, Bielefeld University, D-33501 Bielefeld, Germany}
\author{Stefan Bl\"ugel}
\affiliation{Peter Gr\"unberg Institut and Institute for Advanced Simulation, Forschungszentrum J\"ulich and JARA, 52425 J\"ulich, Germany}

\date{\today}

\begin{abstract}
Quasiparticle spectra of potentially half-metallic Co$_2$MnSi and Co$_2$FeSi Heusler compounds have been calculated within the one-shot $GW$ approximation in an all-electron framework without adjustable parameters. For Co$_2$FeSi the many-body corrections are crucial: a pseudogap opens and good agreement of the magnetic moment with experiment is obtained. Otherwise, however, the changes with respect to the density-functional-theory starting point are moderate. For both cases we find that photoemission and x-ray absorption spectra are well described by the calculations. By comparison with the $GW$ density of states, we conclude that the Kohn-Sham eigenvalue spectrum provides a reasonable approximation for the quasiparticle spectrum of the Heusler compounds considered in this work.
\end{abstract}

\maketitle

\section{Introduction}

Heusler compounds \cite{Heusler} attract ever-growing experimental and theoretical attention, largely because a vast number of such compounds have been predicted to be half-metallic ferromagnets (HMF), i.e., the compounds behave like a metal for one spin channel and like a semiconductor for the other.\cite{Groot83, Kuebler83, Ishida95, Galanakis02} Peculiar electronic transport properties are expected from such materials, e.g., huge magnetoresistive effects in giant and tunnel magnetoresistive (GMR, TMR) devices.

Heusler compounds are ternary  intermetallic compounds with the general chemical formula $X_2YZ$, where $X$ and $Y$ are transition-metal atoms and $Z$ is a main group element. They form the cubic L2$_1$ structure (space group Fm$\bar{3}$m) with a four atom basis. The half-metals among the Heusler compounds follow the Slater-Pauling rule, which connects the magnetic moment per formula unit $m$ and the number of valence electrons $N_\mathrm{V}$ via \cite{Galanakis02}
\begin{equation}
m = N_\mathrm{V} - 24.
\end{equation}

Most theoretical studies of these materials have been based on density functional theory \cite{HK, KS} (DFT) in the Kohn-Sham formalism so far,\cite{Graf11} which gives access to ground-state properties, such as the total energy, atomic forces, magnetic moments etc. It relies on a mapping of the real system onto a fictitious system of noninteracting electrons moving in an effective potential. The half-metallic nature found experimentally for some Heusler compounds is predicted correctly by DFT, together with a quantitative explanation of the Slater-Pauling behavior. However, it is questionable whether the Kohn-Sham eigenvalue spectrum can be taken as the excitation spectrum of the real system. Strictly speaking, there is no theoretical justification for such an interpretation. In fact, while the band structure often resembles the experimentally determined dispersions remarkably well, there are important quantitative discrepancies. For example, the fundamental band gaps of semiconductors and insulators are usually underestimated by a factor of 2 or more. This raises the question if the half-metal band gap is also subject to this underestimation. Studies on Co$_2$MnSi indicate that this is not so: the experimental gap is not larger than about 1\,eV as inferred from tunnel spectroscopy and x-ray absorption experiments.\cite{Kubota09, Sakuraba10, Kallmayer09} This value is very close to the calculated Kohn-Sham gap. The bandwidth of metals and the exchange splitting of ferromagnets are two other important spectral quantities which are often unsatisfactorily described by Kohn-Sham DFT.\cite{Yamasaki03,Schilfgaarde06}

There are several approaches that allow to go beyond Kohn-Sham DFT in this respect. For example, DFT$+U$ and DFT+DMFT (dynamical mean field theory in a correlated subspace) employ an effective, partially screened interaction parameter, the Hubbard $U$ parameter, that acts between electrons in the subspace of localized states while the rest is treated on the level of DFT.\cite{Minar11} The $U$ parameter itself is taken in its static limit. Dynamical screening effects of the itinerant electrons are thus neglected. Furthermore, the Hubbard $U$ parameter is usually taken as an empirical parameter that is fitted to experiment, and the artificial separation into localized and itinerant electrons requires a double-counting correction, which is not uniquely defined. LDA+DMFT calculations on half-metals suggest the presence of nonquasiparticle states inside the half-metal gap, which may destroy the half-metallic character of a material.\cite{Chioncel03, Chioncel08, Chioncel09}

Another method that allows to obtain physical electron addition and removal energies is the $GW$ approximation for the electronic self-energy within many-body perturbation theory.\cite{Hedin65, Aryasetiawan98} In contrast to DFT, the $GW$ method is designed for spectral properties, such as the band structure. Typically, it opens the gap of semiconductors and insulators and gives good agreement with experiments. We apply this method to Co$_2$MnSi and Co$_2$FeSi, two prototypical and potentially half-metallic Heusler compounds, to study the effect of many-body corrections on their band structures. In particular, we will answer the question if the $GW$ approximation increases the half-metal band gap as in the case of semiconductors and insulators or not. As already mentioned above, an increase may worsen the good agreement to experiment achieved by Kohn-Sham DFT.

Co$_2$MnSi and Co$_2$FeSi are particularly interesting because of their large magnetic moments and Curie temperatures. They are known to form the L2$_1$ structure with a low degree of chemical disorder;\cite{Balke06} this allows accurate comparison between experiment and theory. The half-metallic character and integer magnetic moment of Co$_2$MnSi is already predicted by DFT.\cite{Ishida95} For Co$_2$FeSi, DFT calculations predict a significantly reduced magnetic moment with respect to experiment and the Slater-Pauling value.\cite{Balke06} DFT$+U$ and DFT+DMFT find a magnetic moment in accordance with the Slater-Pauling rule and experiment with $U$ parameters of 1.8 and 3\,eV, respectively.\cite{Balke06,Chadov09} However, DFT$+U$ deteriorates the spectral properties of Co$_2$FeSi compared to conventional DFT calculations.\cite{Meinert12} It is the aim of this work to investigate to which extent many-body corrections within the $GW$ method modify or confirm the predictions made by DFT calculations and, in particular, whether the $GW$ approximation is able to rectify the magnetic moment of Co$_2$FeSi without deteriorating the spectral properties.

\section{Method}

In this work we present one-shot $GW$ calculations, which yield the quasiparticle energies $E_{n \mathbf{k}}^\sigma$ as corrections on the Kohn-Sham energies $\epsilon_{n \mathbf{k}}^\sigma$,
\begin{equation}\label{qpe}
E_{n \mathbf{k}}^\sigma = \epsilon_{n \mathbf{k}}^\sigma + \left\langle \phi_{n \mathbf{k}}^\sigma \left| \Sigma^\sigma_\mathrm{xc}(E_{n \mathbf{k}}^\sigma) - v_\mathrm{xc}^\sigma \right| \phi_{n \mathbf{k}}^\sigma \right\rangle,
\end{equation}
where $\phi_{n \mathbf{k}}^\sigma$ are the Kohn-Sham wavefunctions and $n$, $\mathbf{k}$ and $\sigma$ are the band index, Bloch vector, and electron spin, respectively. The quasiparticle correction contains the exchange-correlation potential $v_\mathrm{xc}^\sigma$, for which we employ the PBE functional,\cite{PBE} and the $GW$ self-energy operator, which is given in formal notation by $\Sigma_\mathrm{xc}^\sigma = iG^\sigma W$,\cite{Hedin65} where $G^\sigma$ and $W$ are the Kohn-Sham Green function and screened Coulomb potential, respectively. The latter is approximated by the random-phase approximation $W=v(1-vP)^{-1}$ with the polarization function $P = -i \sum_\sigma G^\sigma G^\sigma$ and the bare Coulomb interaction $v$. Notably, $W$ does not depend on spin: quasiparticles of both spin directions interact via the same screened potential.

We use the \textsc{fleur} \cite{fleur} and \textsc{spex} \cite{Friedrich10} programs for the DFT and $GW$ calculations, respectively. These codes are based on the highly precise all-electron full-potential linearized augmented-plane-wave (FLAPW) method. Transition-metal 3\textit{s}, 3\textit{p} and Si 2\textit{s}, 2\textit{p} semicore states are treated with local orbitals, although their effect on the spectra is small. The muffin-tin radii are set to 2.25 and 2.31\,bohr for the transition-metal atoms and Si, respectively. We employ plane-wave and angular momentum cutoff parameters of $k_\mathrm{max} = 4.0\,\mathrm{bohr}^{-1}$ and $l_\mathrm{max} = 8$. The DFT calculations are performed on 256 $\mathbf{k}$ points in the irreducible wedge to obtain a reliable starting point.

The $GW$ calculations are performed with a $10 \times 10 \times 10$ $\mathbf{k}$-point mesh that contains 47 points in the irreducible wedge with cutoff parameters for the mixed product basis $L_\mathrm{max} = 4$ and $G_\mathrm{max}' = 3.5\,\mathrm{bohr}^{-1}$, and an additional cutoff $\sqrt{4\pi/v_\mathrm{min}} = 4.5\,\mathrm{bohr}^{-1}$ for the correlation part of the self-energy; see Ref.~\onlinecite{Friedrich10} for details. We find that 50 empty bands are sufficient to converge the quasiparticle spectra to better than 0.05\,eV. This is also the estimated accuracy of the $\mathbf{k}$-point sampling. The self-energy is evaluated with a contour integration in the complex frequency plane, and Eq.~\ref{qpe}, which is nonlinear in energy, is solved on an energy mesh with spline interpolation between the points.

The densities of states (DOS) curves are obtained with tetrahedron integration and convoluted with a Gaussian of 0.1\,eV full-width at half-maximum. Binding energies are always taken relative to the corresponding Fermi energy, which is determined by the condition that the DOS integrates to the total number of electrons from $-\infty$ to the Fermi energy. All calculations are based on the experimental lattice constant of 5.64\,\AA{} for both compounds.\cite{Balke06}

\section{Results}

\begin{figure}[t]
\includegraphics[width=8.6cm]{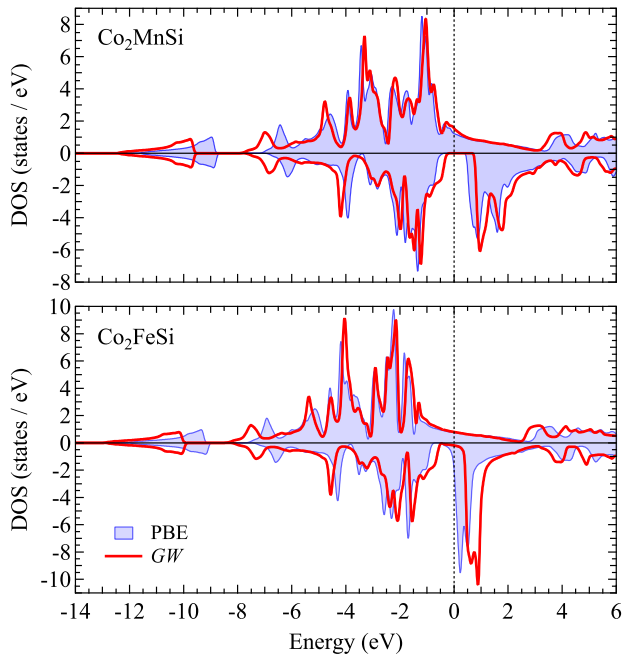}
\caption{\label{Fig1} Kohn-Sham and $GW$ DOS of Co$_2$MnSi and Co$_2$FeSi.}
\end{figure}

In Fig.~\ref{Fig1} we present the PBE and $G W$ DOS of Co$_2$MnSi and Co$_2$FeSi. In both cases, the main effect of the quasiparticle corrections are downshifts of the Si \textit{s} states (between $-9$ and $-12$\,eV) by 0.9\,eV and the hybrid \textit{p-d} states (between $-4$ and $-8$\,eV) by 0.8\,eV-0.5\,eV---see, e.g., Ref.~\onlinecite{Ishida95} for partial DOS plots. Additionally, the exchange splitting of these states is reduced.

\begin{table*}[t]
\caption{\label{Tab1}Magnetic moments (in $\mu_\mathrm{B}$), minority $\Gamma-\Gamma$, minority $\Gamma-X$, and minority spin flip gaps (in eV) of Co$_2$MnSi and Co$_2$FeSi obtained from Kohn-Sham DFT, the $GW$ approximation, and experiment where available.}
\begin{ruledtabular}
\begin{tabular}{l c c c  c c c c c c c}
	&	m$^\mathrm{PBE}$ & m$^{GW}$ &	m$^\mathrm{exp}$ &	$E_{\Gamma \rightarrow \Gamma}^\mathrm{PBE}$  & $E_{\Gamma \rightarrow \Gamma}^{GW}$ & $E_{\Gamma \rightarrow \mathrm{X}}^\mathrm{PBE}$  & $E_{\Gamma \rightarrow \mathrm{X}}^{GW}$ & $E_{\downarrow\uparrow}^\mathrm{PBE}$ & $E_{\downarrow\uparrow}^{GW}$ & $E_{\downarrow\uparrow}^\mathrm{exp}$\\\hline
Co$_2$MnSi	&	5.00	&	5.00	&	4.97\footnotemark[1]	&	0.86	&	0.99	&	0.82 &	0.95 	&	0.37	&	0.17	&	0.25\footnotemark[2], 0.35\footnotemark[3]	\\
Co$_2$FeSi	&	5.52	&	5.89	&	5.97\footnotemark[1]	&	0.94	&	0.92	&	-- &	-- 	&	--	&	--	&	--\footnotemark[2]$^,$\footnotemark[4]	\\
\end{tabular}
\end{ruledtabular}
\footnotetext[1]{Reference \onlinecite{Balke06}}
\footnotetext[2]{Reference \onlinecite{Kubota09}}
\footnotetext[3]{Reference \onlinecite{Sakuraba10}}
\footnotetext[4]{Reference \onlinecite{Oogane09}}
\end{table*}

The binding energies of the occupied \textit{d} states of Co$_2$MnSi remain largely unchanged. While the absolute values of the 3\textit{d} quasiparticle energies do change due to the exactly cancelled self-interaction error, the Fermi energy changes likewise so that the difference remains more or less the same. Close to the Fermi energy we find a small increase of the exchange splitting by 0.2\,eV in Co$_2$MnSi, which places the $GW$ Fermi energy closer to the minority valence band minimum. In addition, the minority gap (given by the $\Gamma \rightarrow \mathrm{X}$ transition) is slightly enhanced from 0.82\,eV to 0.95\,eV, and the unoccupied minority \textit{d} states are rigidly pushed up in energy. The only small quasiparticle correction of the minority gap is noteworthy in view of the fact that semiconductor and insulator gaps usually increase considerably (and rightly so) when treated within the $GW$ approximation. Thus, the apprehension that $GW$ might worsen the agreement with experiment is proved wrong with this result. This aspect will be analyzed in more detail in the next section. 

A similar effect is encountered for Co$_2$FeSi, but the increase of the exchange splitting and the shift of the unoccupied \textit{d} states are larger than in Co$_2$MnSi. This places the Fermi energy in the middle of a minority pseudogap, which accommodates a light band of Fe $t_{2g}$ character.

\begin{figure}[t]
\includegraphics[width=8.6cm]{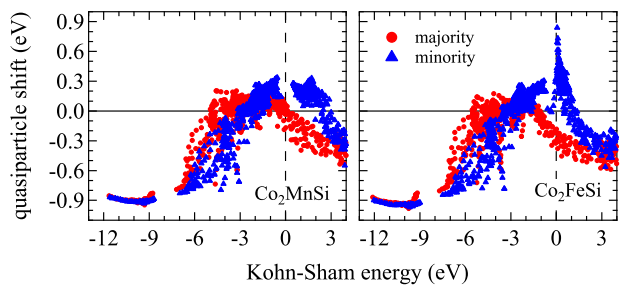}
\caption{\label{Fig2} Quasiparticle shifts as function of the Kohn-Sham energy.}
\end{figure}

The quasiparticle shifts are displayed in Fig.~\ref{Fig2}. We see that the \textit{d} states are pushed up in energy; the occupied states move closer to $E_\mathrm{F}$ and the unoccupied states away from it. Also the increase of the exchange splitting around the Fermi energy becomes visible. For Co$_2$FeSi, the states close to the Kohn-Sham Fermi energy are pushed up in energy by as much as 0.85\,eV. These are mostly of Fe \textit{d} character with 25 - 50\,\% admixture of Co \textit{d} character.

Table~\ref{Tab1} compares the magnetic moments, the minority $\Gamma \rightarrow \Gamma$ and $\Gamma \rightarrow \mathrm{X}$ transition energies, and the minority spin flip gaps from the Kohn-Sham and quasiparticle calculations and from experiments. The magnetic moment of Co$_2$MnSi is the same in PBE and $GW$ and matches the experimental value very well.\cite{Balke06} With the Fermi energy located in the pseudogap, the magnetic moment of Co$_2$FeSi is increased from 5.52\,$\mu_\mathrm{B}$\,/\,f.u. to 5.89\,$\mu_\mathrm{B}$\,/\,f.u., improving the agreement with the experimental value of 5.97\,$\mu_\mathrm{B}$\,/\,f.u. considerably.\cite{Balke06} Hence, the one-shot $GW$ approach manages to correct the magnetic moment. We note that the orbital magnetic moment \cite{Chadov09,Meinert12} is not taken into account in our calculations.

The minority spin flip gap, i.e., the energy required to promote an electron from the minority valence band maximum to a majority state at the Fermi energy, is nonzero for Co$_2$MnSi but zero for Co$_2$FeSi due to the minority pseudogap. From tunnel spectroscopy of magnetic tunnel junctions one deduces a spin flip gap for Co$_2$MnSi between 0.25 and 0.35\,eV.\cite{Kubota09, Sakuraba10} Both theoretical values are in fair agreement with these experimental numbers. Co$_2$FeSi does not have a spin flip gap in the experiment,\cite{Kubota09, Oogane09} which agrees with both calculations. The minority $\Gamma \rightarrow \Gamma$ transition energy is increased for Co$_2$MnSi by 0.13\,eV, whereas it essentially remains the same in the case of Co$_2$FeSi. This is very different from the DFT$+U$ ($U=1.8$\,eV) result, where the $\Gamma \rightarrow \Gamma$ gap of Co$_2$FeSi increases to 1.8\,eV.\cite{Balke06}

\begin{figure}[b]
\includegraphics[width=8.6cm]{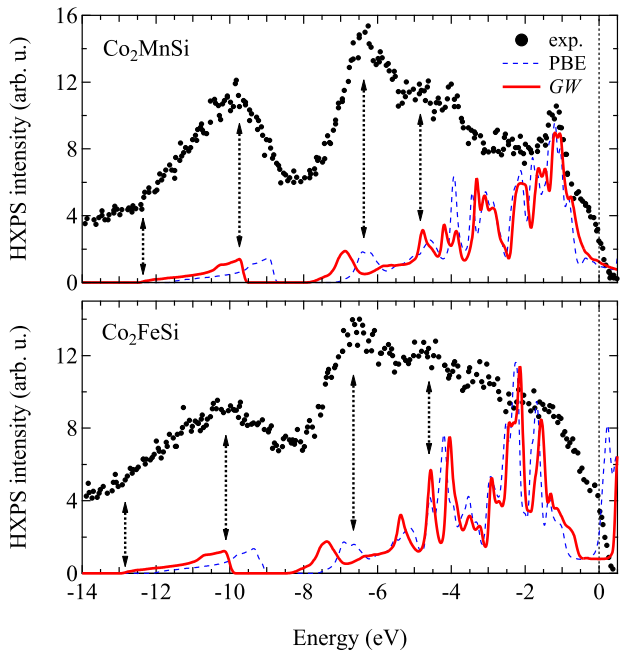}
\caption{\label{Fig3} Comparison of experimental high-energy x-ray photoemission spectra and total DOS of Co$_2$MnSi and Co$_2$FeSi. Experimental data taken from Ref.~\onlinecite{Fecher07}.}
\end{figure}

We compare our calculated quasiparticle spectra with experimental high energy x-ray photoemission spectra (HXPS) taken at 7.935\,keV.\cite{Fecher07} The full valence band spectra are given in Fig.~\ref{Fig3}, with the features discussed in the following marked by arrows. We compare only peak positions, as a detailed analysis of the peak heights would require the calculation of the transition matrix elements, which is beyond the scope of this paper. For both materials, the main features of the spectra are reproduced by the calculations. The valence band minima of Co$_2$MnSi and Co$_2$FeSi at $-12.4$\,eV and $-12.8$\,eV, respectively, are accurately reproduced by the $GW$ calculations. Also the maxima of the emission from the Si \textit{s} states are about correct. The emission maxima of the \textit{p-d} hybrid states are in good agreement with the PBE calculation, whereas $GW$ places them too low in energy compared to experiment, while their onset is described better. It is difficult to assign the individual structures between $-5$\,eV and $E_\mathrm{F}$ in the experimental spectra to the various peaks in the quasiparticle spectra. However, the overall agreement seems to be reasonable in both cases. The plasmon frequency calculated within the random-phase approximation amounts to 4.7\,eV and 6.0\,eV for Co$_2$MnSi and Co$_2$FeSi, respectively, in agreement with previous calculations.\cite{Picozzi06, Kumar09} These energies are well within the valence band region, indicating that the measured x-ray photoemission spectra might be affected by plasmon satellites.

\begin{figure}[t]
\includegraphics[width=8.6cm]{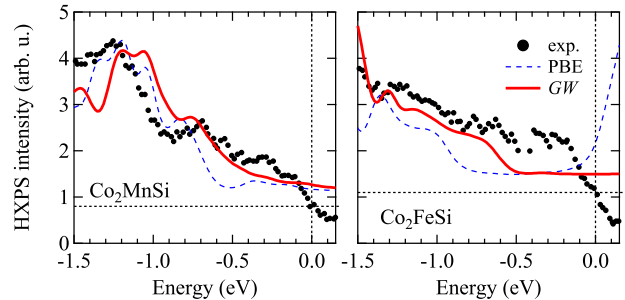}
\caption{\label{Fig4} Comparison of experimental high-energy x-ray photoemission spectra and total DOS of Co$_2$MnSi and Co$_2$FeSi close to the Fermi energy. The horizontal dashed line denotes the additional background added to the theoretical spectra. Experimental data taken from Ref.~\onlinecite{Fecher07}.}
\end{figure}

Additional high-resolution HXPS spectra taken close to the Fermi energy are shown in Fig.~\ref{Fig4}. Both spectra are well described by the $GW$ calculation. For Co$_2$MnSi, the main feature at $-1.25$\,eV, arising from a Co-Mn majority \textit{d} state, is placed 0.1\,eV too high in $GW$. The shoulder at $-0.7$\,eV arises from a pure Co minority \textit{d} state and is reproduced by $GW$. The structure at $-0.3$\,eV in the experimental Co$_2$MnSi spectrum might be related to the minority valence band maximum, which appears at about the same energy in PBE and $GW$, see the spin-flip gap values in Table \ref{Tab1}. Strangely, the $GW$ DOS does not show a structure at this energy in contrast to the PBE DOS. A comparison with Fig.~\ref{Fig1} reveals that while the minority DOS drops at $-0.3$\,eV, the majority DOS happens to increase at exactly the same energy so as to compensate the decrease from the minority states. However, we note that even a small difference in the transition matrix elements of spin-up and spin-down states, which have been neglected in the present work, are expected to produce a structure in the $GW$ spectrum at the correct energy. 

The photoemission spectrum of Co$_2$FeSi close to the Fermi energy in Fig.~\ref{Fig4} is well described by the $GW$ calculation and improves on the PBE result. The features are less pronounced than for Co$_2$MnSi; however, the shoulder at $-1.3$\,eV and the shape of the spectrum below $-0.6$\,eV are reproduced.

\begin{figure}[b]
\includegraphics[width=8.6cm]{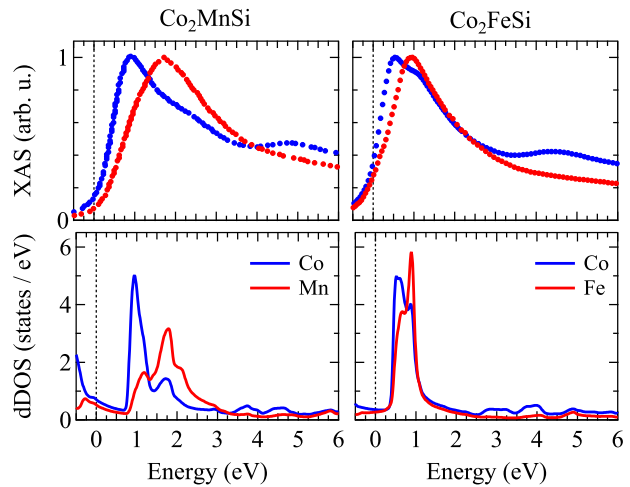}
\caption{\label{Fig5} Top row: experimental spin-averaged x-ray absorption spectra at the Co, Fe, and Mn L$_3$ absorption edges. Bottom row: site-resolved $GW$ \textit{d} electron DOS. The absorption maxima are aligned with the theoretical DOS maxima. Experimental data taken from Refs. \onlinecite{Telling08} and \onlinecite{Meinert12}.}
\end{figure}

Now we turn to the unoccupied states. We focus on the transition-metal \textit{d} states, which can be mapped out element-specifically by soft x-ray absorption spectroscopy at the L$_3$ edges (using the $2p\rightarrow3d$ transitions). In Fig.~\ref{Fig5} we compare the experimental L$_3$ absorption spectra of Co, Mn, and Fe in Co$_2$MnSi and Co$_2$FeSi with the corresponding $GW$ \textit{d} electron DOS. The absorption maxima are aligned with the DOS maxima. The shapes of the spectra agree with the computed DOS; also, the alignments with the Fermi energy seem reasonable, and the hybridizations are visible in spite of the large lifetime broadening of the spectra. For a detailed comparison of the energy levels one would have to take into account the interaction of the core hole with the photoelectron, i.e., an exciton. This effect is of the order of $0.3\dots0.5$\,eV, and it affects the final states in dependence on their symmetry and localization.\cite{Meinert11, Kallmayer09} A consistent treatment of the optical absorption process would require solving the Bethe-Salpeter equation.\cite{Laskowski10} Kallmayer~\textit{et al.}~have taken the exciton binding energy as 0.5\,eV and assumed the exciton to effect a rigid shift of the unoccupied \textit{d} states towards the Fermi level. With these assumptions, they find that the maximum DOS of Co should be at 0.9\,eV and 0.6\,eV above $E_\mathrm{F}$ for Co$_2$MnSi and Co$_2$FeSi, respectively.\cite{Kallmayer09} These values agree with our calculated $GW$ values within 0.1\,eV, while the Kohn-Sham spectrum shows a larger discrepancy, see Fig.~\ref{Fig1}.

The unoccupied minority $d$ states of Co$_2$FeSi are mostly shifted rigidly upwards in the $GW$ calculation. It was recently shown, that x-ray magnetic linear dichroism spectra of Co$_2$FeSi can be described by a DFT calculation with the PBE functional plus a rigid shift of the $d$ states.\cite{Meinert12} We conclude that the spectrum of unoccupied states is described correctly within the $GW$ approximation.

\section{Role of the screening}

In the $GW$ approximation, the screened Coulomb interaction $W(\mathbf{r},\mathbf{r}';\epsilon)$ is the key ingredient. Intuitively, one may expect that the similarity of the PBE and $GW$ results arises from the metallic screening of the majority spin channel. To test this conjecture, we have computed the $GW$ gap of Co$_2$MnSi without metallic screening. We also analyze the importance of local-field effects and briefly discuss results from a one-shot PBE0 hybrid functional scheme.\cite{Ernzerhof99,Perdew96}

Neglecting screening altogether, i.e., replacing $W$ by the bare Coulomb interaction $v$, we obtain the (non-selfconsistent) Hartree-Fock gap of 9.65 eV, a gross overestimation. Now we allow for screening effects but suppress the metallic screening. We achieve this by replacing polarization contributions from the majority spin channel, where metallic screening takes place, by the polarization arising from the minority spin electrons, i.e, we use $P = 2 P_\downarrow$. This enforces a long-range $W$ also in the static limit, since the electrons cannot flow freely in the gapped minority channel, which would enable them to screen test charges completely. Employing this artificial semiconductor-like polarization, which exhibits a finite dielectric constant of $\varepsilon_\infty=14$, we obtain only a slightly larger minority energy gap of 0.97\,eV. On the other hand, setting $P = 2 P_\uparrow$ reduces the gap to 0.86\,eV. Clearly, the majority electrons generate a more effective screening, but the differences in the gap values are relatively small. Long-range metallic screening does not seem to contribute significantly to the total screening, and screening taking place at short distances seems to be more effective.

To investigate this further, we exclude local-field effects. Local-field effects arise from density fluctuations of a different wave length than their generating fields. These couplings are related to the offdiagonal elements of the polarization matrix $P$ represented in a plane-wave basis. (We employ, instead, a basis of eigenvectors of the Coulomb matrix represented in the mixed product basis, which are, however, reasonably close to plane waves.) Setting these offdiagonal elements to zero implies that the screened interaction $W(\mathbf{r},\mathbf{r}';\epsilon)$ only depends on the difference $\left|\mathbf{r} - \mathbf{r}'\right|$ rather than on the absolute positions $\mathbf{r}$ and $\mathbf{r}'$. This is equivalent to saying that the charge density within the unit cell and its screening are homogeneous.\cite{Hybertsen86} The resulting energy gap of 1.65\,eV is nearly twice as large as the Kohn-Sham value. Also, the low-lying \textit{s} and \textit{p-d} states are affected significantly: they shift by about 0.5\,eV upwards in energy with respect to the PBE result, at odds with experiment. Furthermore, the exchange splitting of the occupied \textit{d} states increases and the minority spin-flip gap vanishes. Thus, the charge inhomogeneity plays a crucial role for the screening properties. We note, that Damewood and Fong found similarly small changes of the half-metallic gaps of zincblende CrAs, MnAs, and MnC in the $GW$ approximation with respect to PBE calculations, and a similar behaviour of the gap size with respect to the local-field effects.\cite{Damewood11}

In recent years, potentials derived from hybrid functionals, e.g., PBE0,\cite{Ernzerhof99} have often been used as an approximation to the electronic self-energy. Being nonlocal they fulfill an important condition of the self-energy. Hybrid functionals have been shown to overcome the typical underestimation of band gaps within Kohn-Sham DFT. However, dynamical effects are not taken into account, and screening is only considered in an average way by the parameter that mixes the nonlocal and local parts. In the PBE0 functional this mixing parameter is universally taken to be 0.25.\cite{Perdew96} Since the $GW$ approximation contains the bare exchange exactly, we can easily calculate a one-shot (non-selfconsistent) PBE0 energy spectrum. We find that while PBE0 gives similar results for the binding energies of the low-lying \textit{s} and \textit{p-d} states as the $GW$ approximation, it completely fails in determining the minority gap, for which it yields 3.03\,eV. Also, the exchange splitting is strongly overestimated in both cases. Thus, only a dynamical self-energy can simultaneously describe states close to the Fermi energy and far away equally well.

\section{Conclusions}

We have presented one-shot $GW$ calculations of the (potentially) half-metallic Heusler compounds Co$_2$MnSi and Co$_2$FeSi. The $GW$ quasiparticle spectra are qualitatively similar to the Kohn-Sham eigenvalue spectra, but show important quantitative differences. In particular, the $GW$ approximation predicts an electronic structure with a minority pseudogap in the case of Co$_2$FeSi, which corrects the magnetic moment per unit cell to nearly an integral number, consistent with available experimental data. 

The quasiparticle spectra are in good agreement with photoemission and x-ray absorption data for both compounds. The electronic screening is effective at short distances and charge inhomogeneities play an important role for the screening. Furthermore, it has been shown that the PBE0 hybrid potential cannot be used as an approximate self-energy: it even yields worse results than the local PBE potential.

So far, most theoretical studies of Heusler compounds have been based on the Kohn-Sham band structure. In this work, we have demonstrated that it can, in fact, represent a reasonable approximation to the many-body quasiparticle spectrum, which confirms previous successful calculations of spectral properties of Heusler compounds within DFT.

\acknowledgments
Financial support by the Deutsche Forschungsgemeinschaft (DFG) is acknowledged. We are grateful for helpful discussions with Ersoy \c{S}a\c{s}io\u{g}lu and Jan Schmalhorst.

\end{document}